# The Detection of Thoracic Abnormalities ChestX-Det10 Challenge Results

Jie Lian[1], Jingyu Liu[1], Yizhou Yu[1], Mengyuan Ding[2], Yaoci Lu[3],
Yi Lu[4], Jie Cai[4], Deshou Lin[5], Miao Zhang[4], Zhe Wang[6], Kai He[7]
and Yijie Yu[8]

[1]*Deepwise AI Lab*
[2]*Nanjing University of Science and Technology*
[3]*Xinhua College of Sun Yat-sen University*
[4]*Zhejiang University*
[5]*Zhejiang Sci-Tech University*
[6]*Shaanxi Normal University*
[7]*University Of Electronic Science And Technology Of China*
[8]*Jiangsu University*

## 1 Introduction

The detection of thoracic abnormalities challenge is organized by the Deepwise AI Lab. The challenge is divided into two rounds. In this paper, we present the results of 6 teams which reach the second round. The challenge adopts the ChestX-Det10 [1] dateset proposed by the Deepwise AI Lab. ChestX-Det10 is the first chest X-Ray dataset with instance-level annotations, including 10 categories of disease/abnormality of 3,543 images. The annotations are located at https://github.com/Deepwise-AILab/ChestX-Det10-Dataset. In the challenge, we randomly split all data into 3001 images for training and 542 images for testing.

## 2 ChestX-Det10 Challenge Results.

In this section, we report the results of the second round of ChestX-Det10 challenge. For practical usage, we use recall (sensitivity) at fixed false positives per image as the main criteria. The threshold of positive sample IOU is set to 0.5. In particular, we set the rate of instance-level FP/image to be 0.05, 0.1 and 0.2. Finally, we take the average of these three values. Moreover, we report the average predict time per image on 542 samples (ChestX-Det10 test-set). All models are test on single GPU (GeForce GTX 1080 Ti) and batch size is set to 1. A total of 6 teams are compared in the second round. We rank all teams according to the average recall. Table 1 shows detailed detection results.

| No. | Recall | | | | time (s/image) |
|---|---|---|---|---|---|
| | @0.05fp/image | @0.1fp/image | @0.2fp/image | average | |
| 1 | 0.4405 | 0.5108 | 0.5772 | 0.5095 | 0.8173 |
| 2 | 0.4303 | 0.5051 | 0.5666 | 0.5007 | 0.8432 |
| 3 | 0.4152 | 0.4869 | 0.5675 | 0.4899 | 0.9613 |
| 4 | 0.4213 | 0.4918 | 0.5542 | 0.4891 | 0.9576 |
| 5 | 0.4144 | 0.4814 | 0.5516 | 0.4824 | 0.9613 |
| 6 | 0.4007 | 0.4752 | 0.5358 | 0.4706 | 0.8063 |

Table 1: Detection results of each team in the second round of ChestX-Det10 challenge.

## 3 Submitted Detection Schemes.

In this section, we provide a short summary of all submitted detection schemes that are considered in the second round of ChestX-Det10 challenge.

### No.1

*Mengyuan Ding, Yaoci Lu*

**Method.** Using TSD [2] as the baseline and the backbone employs ResNeXt101 64×4d [3].

**Anchor.** There exists the situation that one box completely contains another smaller one. It may cause harder feature representation. Therefore, using fine-grained anchors to match each ground truth is more reasonable. Anchor scales are set to [2, 4, 8, 12] and aspect ratios are set to [0.5, 0.75, 1.0, 1.5, 2.0, 3.0].

**Data-augment.** The data augmentation adopts the method in [4]. First, randomly choose a ratio in [r1, r2] and a ground truth in the image. Then, crop a patch from the image according to the ratio and the centre point of the ground truth. Finally, resize the patch to the cropped size which is set in advance and adjust the corresponding ground truth. This method has some advantages. It is similar to RandomCrop. But it zooms all cropped images to generate more ground truths with different scales, so it makes training datasets more diversity. Moreover, it equals to the method of multi-scale training and testing. But multi-scale training resizes the whole image and does not change the relative ratio of box-image. Besides, multi-scale testing will increase the inference time greatly. During the training, the cropped size 640*640 is taken as the input size. The image size is adjusted to 820*820 during the testing.

**TSD.** Task-aware spatial disentanglement (TSD) is proposed by [2] to resolve the spatial misalignment between classification and localization in the sibling head. TSD decouples classification and regression from the spatial dimension by generating two disentangled proposals.

# No.2

*Yi Lu, Jie Cai*

**Method.** Using Cascade RCNN as the baseline and the backbone employs ResNeXt101 32×4d. Moreover, the BFN structure which is presented in Libra RCNN and DCN-v2 are also adopted to improve the detection performance. The input resolution of image is fixed at 1152×1152 at both training and testing procedure. As for augmentation method, random horizontal flip and gridmask are used at training stage. During testing stage, only horizontal flip is retained. The SGDM with gradient centralization is used as the optimization method to speed up the convergence procedure.

# No.3

*Deshou Lin*

**Method.** Using Cascade RCNN as the basline. The DCN and FPN are applied at backbone ResNext101, which enhances the transformation modeling capability of CNNs and improves the accuracy of the small objects. Besides, some tricks are also adopted to improve the accuracy of results. Firstly, the threshold of positive sample IOU is changed to 0.6 at RPN stage and rcnn stage. Besides, using 0.4, 0.5, 0.6 as the positive sample IOU in cascade heads respectively. Second, keep the original aspect ratio of the images and adopt the multi-scale strategy for both training (shorter side=800 ∼ 1400) and testing (shorter side= 960, 1024, 1088). Finally, the results from multi-scale testing are mixed by WBF (Weighted-Boxes-Fusion).

## No.4

*Miao Zhang*
*zhangmiao@zju.edu.cn*

**Method.** Using Cascade RCNN as the basline. The backbone is X101-64 combined with FPN and DCN-v2. For data augmentation, horizontal flip, multi-scale sampling and gridmask are used to resist overfitting. Besides, smoothl1 loss is replaced by l1 loss to improve performance.

## No.5

*Zhe Wang, Kai He*

**Method.** Using TSD [2] as the baseline and the backbone employs ResNeXt101 64×4d [3]. The weights are initialized from pretrained model on COCO dataset [5]. Similar to [6], k-means clustering is used to determine bounding box priors. Anchor scales are set to [5.5, 6.3, 8.6] and anchor ratios are set to [0.64, 1.27, 2.87]. IoU-balanced Sampling [7] is employed to reduce imbalance between positives and negatives and Balanced L1 Loss [7] is used to trade off between classification loss and location loss. The performance is further improved by performing training and inference on multi-scale inputs. During training, the length of the short side ranges from 600 to 1200. During inference, the length of the short side is randomly selected between 800 and 1000. Notice that the aspect ratio of the inputs always keep unchanged.

## No.6

*Yijie Yu*

**Method.** Using Cascade R-CNN as the baseline. The DCN and FPN are applied at backbone. Multi-scale training [1000, 1200, 1400], cutout, flip, histogram equalization and brightness transform are used for data augmentation. Considering the long tail distribution of data, weighted for different categories of losses.

# References

[1] Jingyu Liu, Jie Lian, and Yizhou Yu. Chestx-Det10: Chest x-ray database on detection of thoracic abnormalities. *arXiv preprint arXiv:2006.10550v3*, 2020.

[2] Guanglu Song, Yu Liu, and Xiaogang Wang. Revisiting the sibling head in object detector. *In CVPR*, 2020.

[3] S. Xie, R. Girshick, P. Dollar, Z. Tu, and K. He. Aggregated residual transformations for deep neural networks. In *Proceedings of the IEEE Conference on Computer Vision and Pattern Recognition*, pages 1492–1500, 2017.

[4] Liu W, Liao S, Hu W, Liang X, and Chen X. Learning efficient single-stage pedestrian detectors by asymptotic localization fitting. *In ECCV*, 2018.

[5] Tsung-Yi Lin, Michael Maire, Serge Belongie, James Hays, Pietro Perona, Deva Ramanan, Piotr Dollar, and C Lawrence tZitnick. Microsoft COCO: Common objects in context. In *European conference on computer vision*, pages 740–755, 2014.

[6] Redmon and A. Farhadi. YOLO-v3: An incremental improvement. *arXiv preprint arXiv:1804.02767*, 2018.

[7] Pang, K. Chen, J. Shi, H. Feng, W. Ouyang, and D. Lin. Libra R-CNN: Towards balanced learning for object detection. *In CVPR*, 2019.